\documentclass[a4paper, conference]{IEEEtran}

\ifCLASSINFOpdf
\else
\fi

\usepackage{cite}
\usepackage{amsmath}
\usepackage{amssymb}
\usepackage{epsfig}
\usepackage{graphicx}
\usepackage{subfigure}

\begin{document}
%
\title{A Novel Compact Tri-Band Antenna Design for WiMAX, WLAN and Bluetooth Applications}

\author{\IEEEauthorblockN{Peshal B. Nayak, Sudhanshu Verma and Preetam Kumar}
\IEEEauthorblockA{Department of Electrical Engineering\\
Indian Institute of Technology Patna\\
Patna, Bihar: 800013\\
Email: nayak.ee10@iitp.ac.in; sverma@iitp.ac.in; pkumar@iitp.ac.in}
}


%


\maketitle

\begin{abstract}
A novel and compact tri-band planar antenna for
2.4/5.2/5.8-GHz wireless local area network (WLAN), 2.3/3.5/5.5-GHz Worldwide Interoperability for Microwave Access (WiMAX) and Bluetooth applications is proposed and studied in this paper. The antenna comprises of a L-shaped element which is coupled with a ground shorted parasitic resonator to generate three resonant modes for tri-band operation. The L-shaped element which is placed on top of the substrate is fed by a 50$\Omega$ microstrip feed line and is responsible for the generation of a wide band at 5.5 GHz. The parasitic resonator is placed on the other side of the substrate and is directly connected to the ground plane. The presence of the parasitic resonator gives rise to two additional resonant bands at 2.3 GHz and 3.5 GHz. Thus, together the two elements generate three resonant bands to cover WLAN, WiMAX and Bluetooth bands of operation. A thorough parametric study has been performed on the antenna and it has been found that the three bands can be tuned by varying certain dimensions of the antenna. Hence, the same design can be used for frequencies in adjacent bands as well with minor changes in its dimensions. Important antenna parameters such as return loss, radiation pattern and peak gains in the operating bands have been studied in detail to prove that the proposed design is a promising candidate for the aforementioned wireless technologies.\footnote{This document is an author's version of \cite{nayak2014novel}.}  
\end{abstract}



%
\IEEEpeerreviewmaketitle

\section{Introduction}
In recent times, there has been a rapid growth in wireless communication systems due to an ever increasing market demand for products that are capable of providing multiple services within a single small device. Also, since FCC (Federal Communications Commission) allowed potential users to make an unlicensed use of medical, industrial and scientific frequencies, the scientific community has seen a great opportunity to design wireless devices that would communicate over short distances. Common examples are the Bluetooth which operates at the internationally available ISM band at 2.4 GHz or the wireless local area network (WLAN) which operates at 2.4, 5.2 and 5.8 GHz. Other wireless technologies such as Worldwide Interoperability for Microwave Access (WiMAX) are also playing an important role in our day to day life. \\
\indent
WLAN implemented as an alternative for or an extension to wired LAN is a flexible data-communications system \cite{ nayak2017multi, nayak2019ap, nayak2019modeling, nayak2019virtual, nayak2016performance, peshal2019modeling}. WLANs make use of radio frequency technology and transmit and receive data over the air. This minimizes the need for wired connections and combines connectivity with user mobility. Nowadays, WLANs are becoming popular in a number of vertical markets such as retail, healthcare, warehousing, manufacturing and academia which have profited from the use of hand-held terminals for real-time information transmission to centralized hosts for processing. WLANs are also being widely recognized as a reliable, cost effective solution for wireless high speed data connectivity and a general purpose connectivity alternative for a broad range of applications. There are three operation bands in the IEEE 802.11 WLAN standards: 2.4 GHz (2400-2484 MHz), 5.2 GHz (5150-5350 MHz) and 5.8 GHz (5725-5825 MHz). WLANs working at IEEE 802.11a employ the higher frequency band from 5.15-5.35 GHz and 5.725-5.825 GHz while those working at IEEE 802.11b/g use the 2.4-2.484 GHz band. 802.11a is usually found on business networks due to its higher cost.\\
\indent
Bluetooth is a technology by means of which a mobile device gets connected with other devices when the former comes within the range of one of the other devices. This range typically varies from 20 feet to 15 meter and the frequency band used is the 2.4 GHz band (2.4-2.4835 GHz). The Bluetooth specification is controlled by the Bluetooth special interest group and the standard is IEEE 802.15.1. The IEEE 802.16, better known as WiMAX can be termed partially a successor to the WiFi protocol. It not only allows higher data rates over longer distances but also an efficient use of bandwidth with minimum interference. In the Asia Pacific region, frequencies available for WiMAX deployments are in the 2.3/3.3/3.5/5.5 GHz range.\\
\indent
As wireless devices have become an integral part of the lives of most people, the integration of technologies such as WLAN, WiMAX, Bluetooth, etc into a single device is a perfect solution to enhance commercial advancements. Although a wide band or ultrawideband \cite{nayak2012ultrawideband} antenna could be a possible solution, systems with such antennas need additional filters to remove interference from communication systems operating in nearby bands. In such a scenario, a multiband antenna \cite{nayak2013multiband, endluri975low} turns out to be a cost effective solution as it does away with the filters by suppressing dispensable bands and thus helps in integrating multiple wireless communication standards in a single system; effectively improving the portability of a modern personal wireless terminal device. \\
\indent
The planar monopole antenna, because of its attractive characteristics including low profile and weight, low cost, and versatile structure for exciting wide impedance bandwidth, dual- or multiresonance mode, and desirable radiation characteristics, has become a preferred candidate among the known triple/multiband antenna prototypes. However, antenna designers face a great challenge when a size reduction of an antenna has to be achieved while increasing the number of operating frequency bands. \\
\indent
But as compact multiband antennas are important for integrating multiple communication standards in a single system, thus effectively improving the portability of a modern personal wireless terminal device, they have attracted great attention.  A number of innovative antenna designs have been proposed for achieving multiple bands of operation such as the use of meandered T-shape \cite{chang2009meandered}, use of meandering slots \cite{hsieh2009compact}, double-T shaped element \cite{kuo2003printed}, use of inverted FL shapes \cite{nakano2005inverted}, use of shorted parasitic element \cite{wong2005dual,sun2012dual,nayak2013compact} and use of shorted parasitic inverted-L wire \cite{jan2004small}. However, these designs can provide only a dual band operation. Additionally, the slot based antenna in \cite{hsieh2009compact}, though much smaller than traditional monopoles, has a structure which is complex for practical applications. The same is true for \cite{nakano2005inverted,wong2005dual,sun2012dual,nayak2013compact,jan2004small} which make use of complicated structures such as shorting pins thereby adding to the complexity of the design as well as the fabrication cost of the antenna. A simple low cost design with a broadband performance has been proposed in \cite{li2008novel}. This antenna is, however, dual band and also has a complicated geometry. Such large and complicated geometries face a problem when the antenna has to be embedded into a small space as in the case of a compact mobile terminal. \\
\indent
Some novel designs have been developed to achieve a tri-band performance. In \cite{chen2008band}, design of a band reject function has been proposed by inserting the proper strips on a wideband printed open slot antenna. In \cite{jan2005bandwidth}, again a printed slot antenna has been proposed. They have the advantage of having wide impedance bandwidth, low profile, light weight and ease of manufacture. But the disadvantage is that they are large in size which leads to technical problems described previously. \\
\indent
Tuning slot and triangular-slot coupled patch antennas were proposed to generate a multiband operation in \cite{chen2005studies}. However, the antenna structure, being somewhat complicated in nature, will increase cost or complexity for practical terminal design. A similar problem will be faced in \cite{zhu2010compact}, where a monopole antenna uses a metamaterial loading to achieve a triple band performance.\\
\indent 
In this paper, a tri-band antenna with an extremely simple structure and using a highly compact radiator to cover all the 2.4/5.2/5.8 GHz WLAN, 2.3/3.5/5.5 GHz WiMAX and the Bluetooth bands of operation is proposed. The antenna consists of a microstrip fed L-shaped element on top and a ground shorted parasitic resonator placed on the other side to generate three wide bands centred at 2.3, 3.5 and 5.5 GHz respectively. The parasitic resonator being placed on the other side of the L-shaped element, is directly shorted with the ground. Since only a single feed has been used and no shorting pins are present, the design is highly compact and extremely simple. The antenna is designed and studied with the high performance full-wave electromagnetic (EM) field simulator Ansoft HFSS software. The paper has been organised as follows. Section II provides a detailed description of the proposed antenna design. This is followed by the results of a parametric study on the antenna in Section III which shows how the bands can be tuned independently. A detailed discussion on the results has been provided in Section IV followed by comparison with latest antenna designs in Secion V and finally conclusion in Section VI. 

\section{Antenna Design}
\begin{figure}[!]
  \centering
  \subfigure[]
  {
      \includegraphics[width=3.5in]{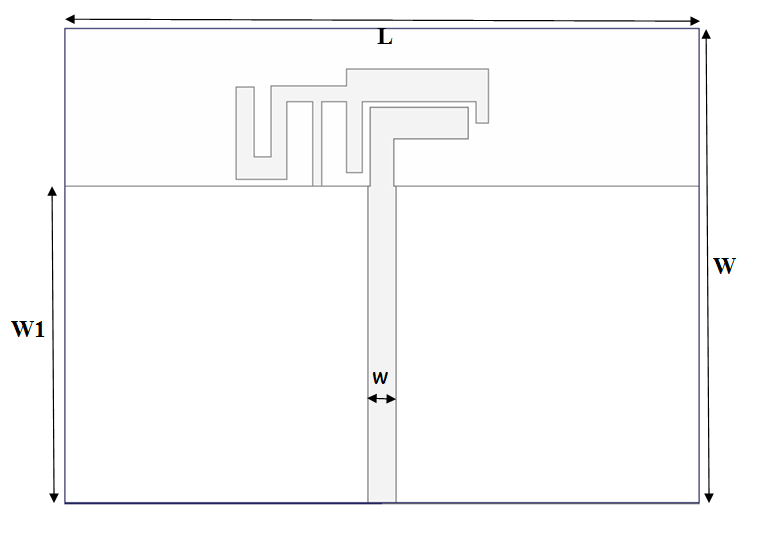}
      \label{fig1a}
  }
  \\
  \subfigure[]
  {
      \includegraphics[width=3.5in]{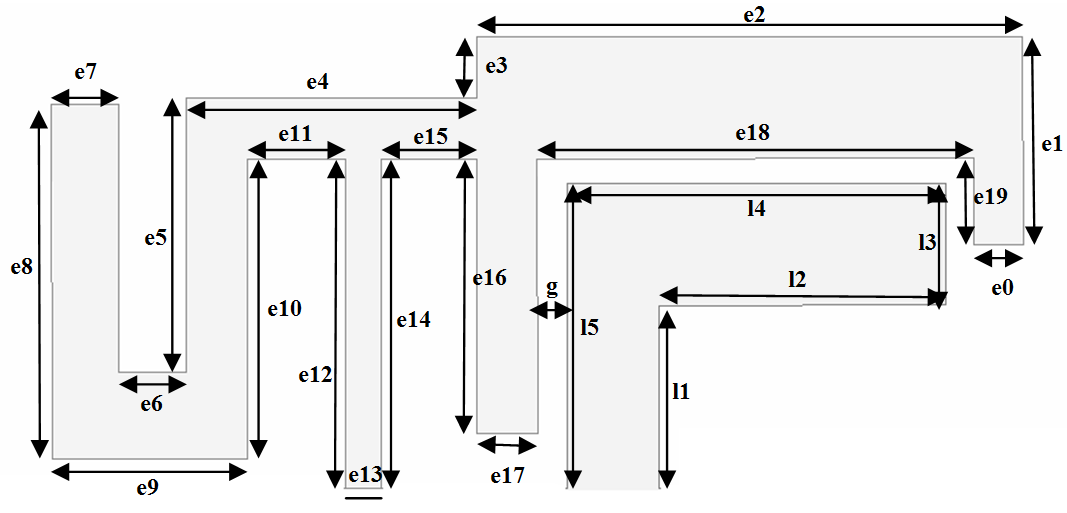}
      \label{fig1b}
  }
    
\caption{(a) Design of the proposed tri-band antenna (b) Layout of the compact radiator}
  \label{fig1}
\end{figure}

The geometry of the proposed antenna has been shown in fig.~\ref{fig1a} and specifications of the design have been given in Table~\ref{table1}. A microstrip feed line of width 1.8 mm has been used in order to achieve a characteristic impedance of 50 $\Omega$. A layout of the radiator has been given in fig.~\ref{fig1b}. The planar antenna consists of an L-shaped patch (element 1) which is fed directly by the microstrip feed line and is placed on the top section of the dielectric substrate and a parasitic resonator (element 2) placed on the bottom and is directly connected to the ground. Element 1 is responsible for generation of the wide 5GHz frequency band for the higher WLAN bands at 5.2 and 5.8 GHz as well as the WiMAX band at 5.5 GHz. The presence of element 2 leads to the generation of the bands at 2.3 GHz and 3.5 GHz for the lower bands of WLAN, WiMAX and Bluetooth. If seen from the top, element 2 seems to surround element 1 with a gap `g' between them. Together they occupy an area of around 15.9 mm x 7.4 mm. The ground plane is around 40 mm x 20 mm in size and the overall dimensions of the antenna are 40 x 30 x 0.8 mm$^{3}$. The antenna has been designed on a dielectric substrate with a relative permittivity of 3.5 and a loss tangent of 0.02. The antenna dimensions have been optimized using computer simulations. The optimized dimensions have been stated in Table~\ref{table2}.

\begin{table}[!]
\caption{Antenna Design Details}
\small\addtolength{\tabcolsep}{0pt}
\centering
\begin{tabular}{|l|c|p{3 cm}|}
\hline
{\bf Material} &  FR$4$\\
\hline
{\bf Dielectric constant} & $3.5$\\
\hline
{\bf Loss Tangent} & $0.02$\\
\hline
{\bf Substrate Thickness} & $0.8$ mm\\
\hline
\end{tabular}
\label{table1}
\end{table}

\begin{table}[!]
\caption{Value of various dimensions in the antenna design (in mm)}
\small\addtolength{\tabcolsep}{0pt}
\centering
\begin{tabular}{|l|l|l|l|l|l|l|l|p{3 cm}|}
\hline
{\bf e0} &  $0.8$ & {\bf e1} & $3.4$ & {\bf e2} & $8.9$ & {\bf e3} & $1.0$ \\
\hline
{\bf e4} & $4.8$ & {\bf e5} & $4.4$ & {\bf e6} & $1.1$ & {\bf e7} & $1.1$\\
\hline
{\bf e8} & $5.8$ & {\bf e9} & $3.2$ & {\bf e10} & $4.9$ & {\bf e11} & $1.6$\\
\hline
{\bf e12} & $5.4$ & {\bf e13} & $0.6$ & {\bf e14} & $5.4$ & {\bf e15} & $1.6$\\
\hline
{\bf e16} & $4.5$ & {\bf e17} & $1.0$ & {\bf e18} & $7.2$ & {\bf e19} & $1.4$\\
\hline
{\bf l1} & $3.0$ & {\bf l2} & $4.7$ & {\bf l3} & $2.0$ & {\bf l4} & $6.2$\\
\hline
{\bf l5} & $5.0$ & {\bf w} & $1.8$ & {\bf g} & $0.5$ &  {\bf W} & $30.0$\\
\hline
{\bf L} & $40.0$ &  {\bf W1} & $20.0$ & & & &\\
\hline

\end{tabular}
\label{table2}
\end{table}

\section{Parametric Study}
\begin{figure}[!]
\centering
\includegraphics[width=3.6 in]{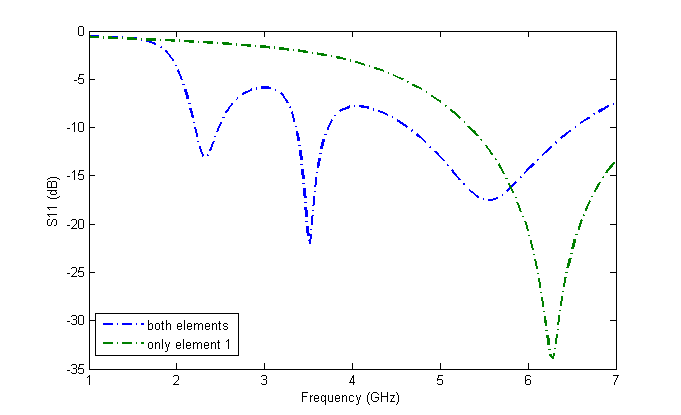}
\caption{Return loss plot for variation in the number of elements}
\label{fig2}
\end{figure}

\begin{figure}[!]
  \centering
  \subfigure[]
  {
      \includegraphics[width=2in]{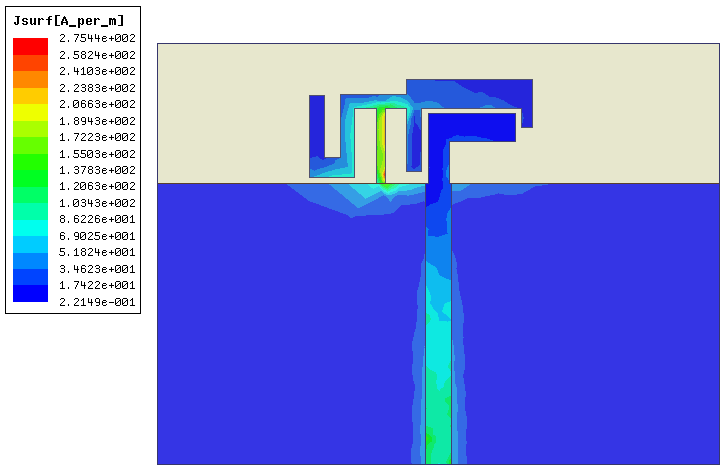}
      \label{fig3a}
  }
  \\
  \subfigure[]
  {
      \includegraphics[width=2in]{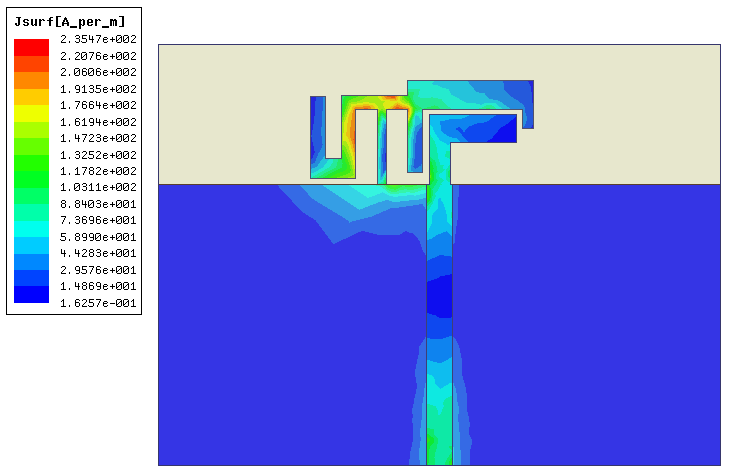}
      \label{fig3b}
  }
  \\
  \subfigure[]
  {
      \includegraphics[width=2in]{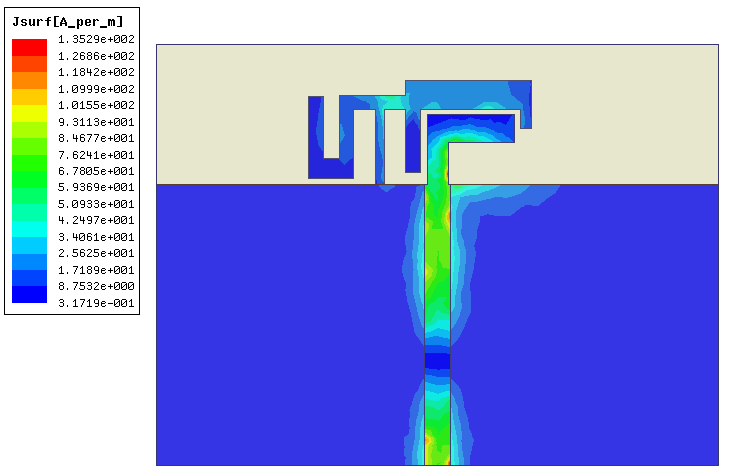}
      \label{fig3c}
  }  
    
\caption{Simulated current distribution at (a) 2.3 GHz (b) 3.5 GHz (c) 5.5 GHz}
  \label{fig3}
\end{figure}

\begin{figure}[!]
\centering
\includegraphics[width=3.6 in]{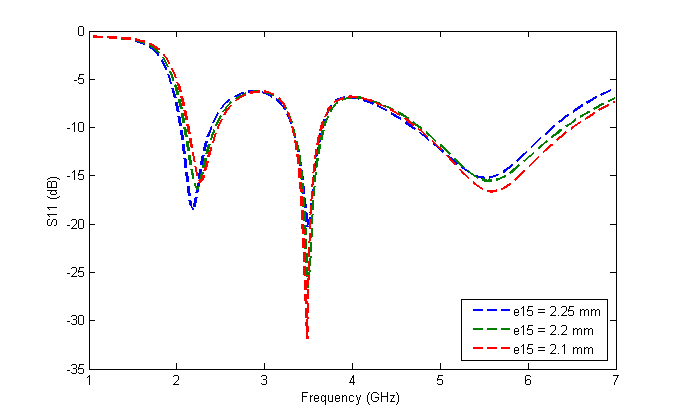}
\caption{Return Loss with variation in the length of e15}
\label{fig4}
\end{figure}

\begin{figure}[!]
\centering
\includegraphics[width=3.6 in]{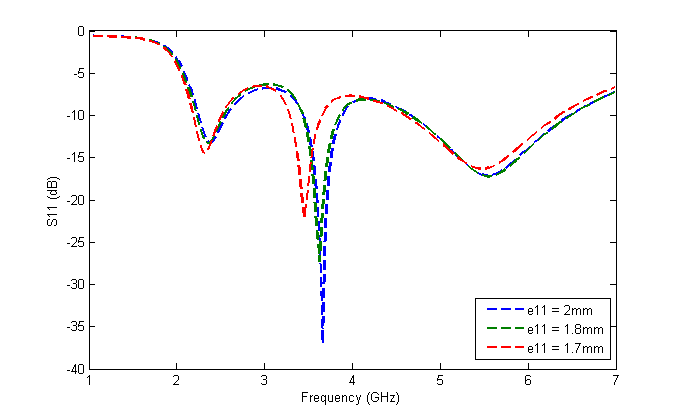}
\caption{Return Loss with variation in the length of e11}
\label{fig5}
\end{figure}

\begin{figure}[!]
\centering
\includegraphics[width=3.6 in]{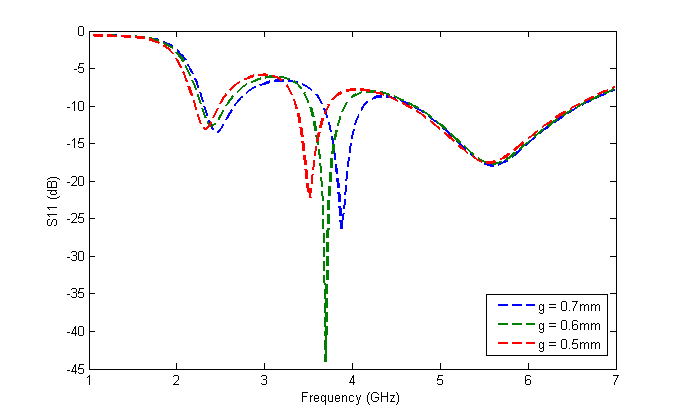}
\caption{Return Loss with variation in the length of g}
\label{fig6}
\end{figure}

\begin{figure}[!]
\centering
\includegraphics[width=3.6 in]{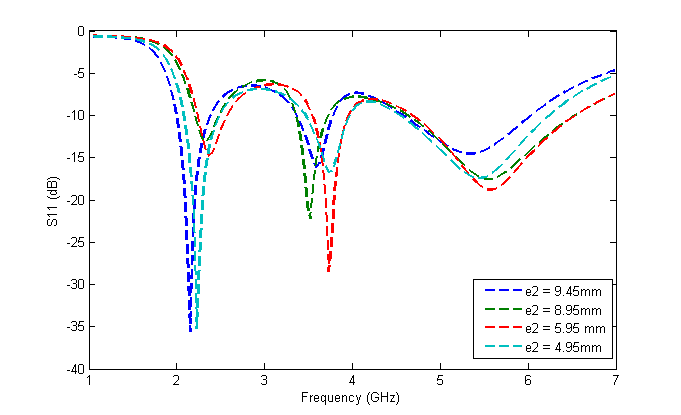}
\caption{Return Loss with variation in the length of e2}
\label{fig7}
\end{figure}

In order to study the effect of various parameters on the resonant frequencies, a detailed parametric study was performed on the antenna. Initially, effect of varying the number of elements was studied. When only element 1 is present, a single wide band centred at around 6.2 GHz is generated as shown in fig.~\ref{fig2}. Upon the insertion of element 2, the higher frequency wideband shifts downwards with its centre frequency becoming 5.5 GHz. In addition to this change, two new bands in the lower frequency range are generated- one around 2.3 GHz and another around 3.5 GHz. Thus, the insertion of element 2 not only affected the higher band but also gave rise to two new bands in the lower frequency range leading to a multiband performance.\\
\indent
The operation of the antenna was further studied by using the current distributions at 2.3, 3.5 and 5.5 GHz. At 2.3 and 3.5 GHz the current was mainly on element 2, though on different sections of it, which contributed to resonance. Fig.~\ref{fig3a} and fig.~\ref{fig3b} show the current distributions at these two operating frequencies. At 5.5 GHz, the current on element 1 was quite large, as shown in fig.~\ref{fig3c}, thus contributing to resonance at the higher frequency.\\
\indent  
Further, the effects of varying various dimensions of the antenna was studied. It was confirmed from this study that the resonant frequencies of the antenna were sensitive to variations in dimensions such as e2, e11, e15 and g. The effect of varying the values of these dimensions is presented in the following paragraphs.\\
\indent
When the value of e15 was varied from 2.1mm to 2.2mm and finally 2.25mm, the resonant frequency of the two bands at 3.5 GHz and 5.5 GHz remained almost constant while the lower band shifts from 2.26 GHz to 2.22 GHz and finally 2.19 GHz. This implies that e15 can be used for fine tuning of the lower band without affecting the two higher bands. Further, as shown in fig.~\ref{fig4}, as the length of this dimension increases the resonant frequency of the lower 2 GHz band of operation decreases.\\
\indent
Next, the value of e11 was changed from 1.7mm to 1.8mm and 2mm. Consequently, the centre frequency of the second band (3.5 GHz) changed from 3.45 GHz to 3.625 GHz and finally to 3.66 GHz as shown in fig.~\ref{fig5}. The second band, thus, shifted rightwards towards higher frequencies as this length was changed. During this change, the first and the third band nearly remained fixed. Thus, this band can be used to fine tune the second band while keeping the first and the third bands nearly unaffected.\\
\indent
Following this, the length of the gap g, was increased and it was noticed as the gap increased from 0.5mm to 0.6mm and 0.7mm, the first and the second band shifted rightwards whereas the third band was almost unaffected as shown in fig.~\ref{fig6}. Thus, this length helps in simultaneous tuning of the two lower bands while maintaining the higher band fixed.\\
\indent 
The length e2 helps in tuning the higher 5 GHz band as well. As this dimension is varied, all the bands shift in a coarse manner as can be seen in fig.~\ref{fig7}. This dimension helps to achieve the tuning of the higher band. However, in this process, the other two bands are also affected.\\
\indent 
From the above parametric study, a general method for designing a tri-band antenna with the same design for applications operating in adjacent frequencies can be realized:
\begin{enumerate}
\item In the first step, the element 1 would have to be inserted while maintaining its dimensions nearly the same as the one in the design. This will create a single resonant band centred at a higher frequency.
\item Then element 2 would have to be inserted such that its dimensions are in the vicinity of the optimized dimensions used for the current design. This will give rise to a tri-band performance in which the lower, the middle and the upper frequency bands will be close to the ones achieved by our optimized design.
\item Following this, the higher frequency band would have to be fine tuned by varying the value of the dimension e2. This will help in achieving the desired higher resonant frequency. However, the other two frequencies will be affected during this process.
\item Then, the values of e15, e11 or g may be adjusted such that the upper two bands also achieve the desired centre frequencies. During this process, the upper band will remained fixed and thus all the three bands will have the expected centre frequencies.
\end{enumerate}

\section{Results and Discussion}
\begin{figure}[!]
\centering
\includegraphics[width=3.6 in]{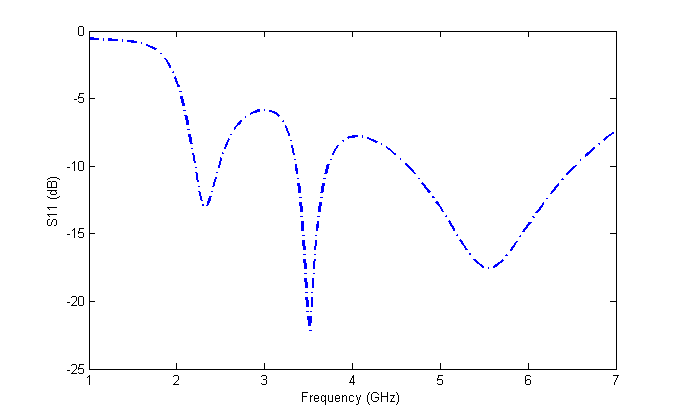}
\caption{Return Loss of the antenna}
\label{fig8}
\end{figure}

\begin{figure*}[!t]
  \centering
  \subfigure[]
  {
      \includegraphics[width=1.4in]{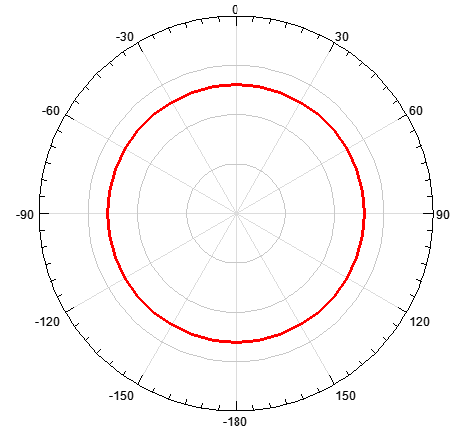}
      \label{fig9a}
  }
  \subfigure[]
  {
      \includegraphics[width=1.4in]{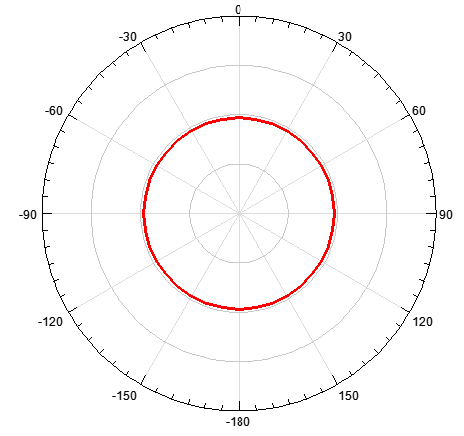}
      \label{fig9b}
  }
  \subfigure[]
  {
      \includegraphics[width=1.4in]{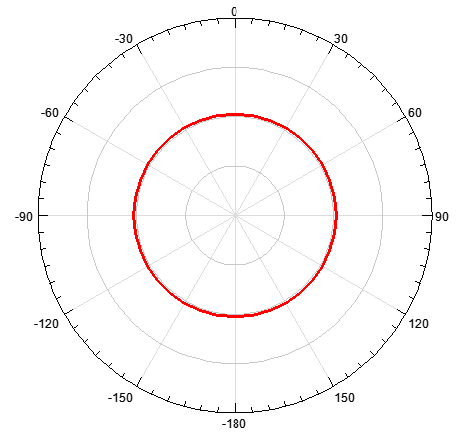}
      \label{fig9c}
  }  
  \\
  \subfigure[]
  {
      \includegraphics[width=1.4in]{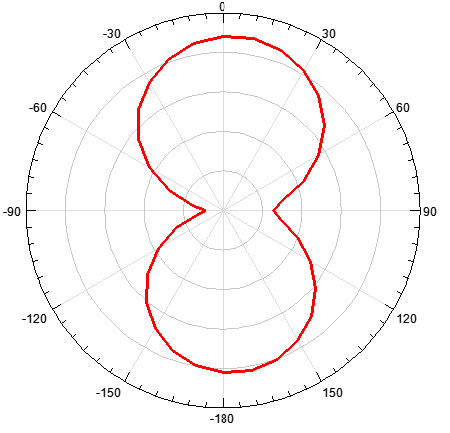}
      \label{fig9d}
  }
  \subfigure[]
  {
      \includegraphics[width=1.4in]{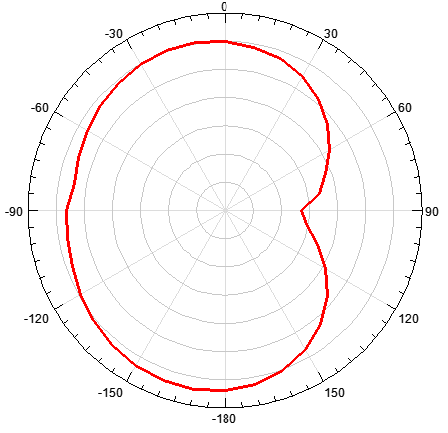}
      \label{fig9e}
  }
  \subfigure[]
  {
      \includegraphics[width=1.4in]{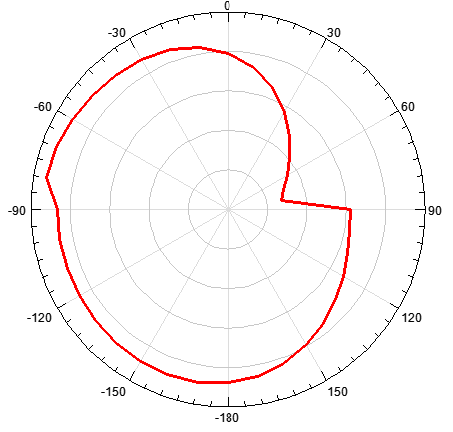}
      \label{fig9f}
  }
  \\
  \subfigure[]
  {
      \includegraphics[width=1.4in]{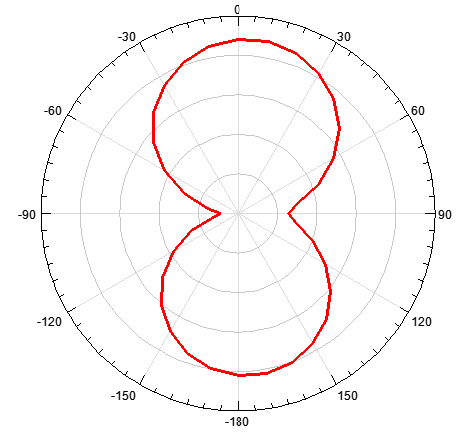}
      \label{fig9g}
  }
  \subfigure[]
  {
      \includegraphics[width=1.4in]{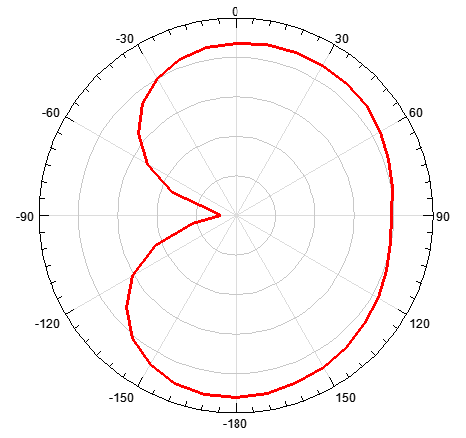}
      \label{fig9h}
  }
  \subfigure[]
  {
      \includegraphics[width=1.4in]{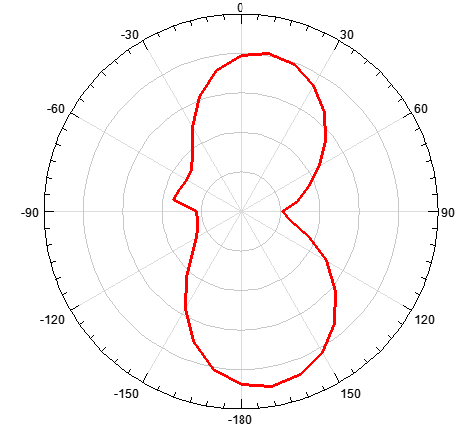}
      \label{fig9i}
  }  
\caption{Radiation pattern in XY plane for (a) 2.3 GHz (b) 3.5 GHz (c) 5.5 GHz;
 Radiation pattern in YZ plane for (c) 2.3 GHz (d) 3.5 GHz (e) 5.5 GHz;
 Radiation pattern in ZX plane for (f) 2.3 GHz (g) 3.5 GHz (h) 5.5 GHz }
  \label{fig9}
\end{figure*}

The return loss plots and the radiation pattern for the antenna have been shown in fig.~\ref{fig8} and fig.~\ref{fig9} respectively. It can be clearly seen that antenna has three bands of operation. The lowest frequency band is centred at 2.3 GHz and has a bandwidth (S11 $<$ -10dB) of 315 MHz (2.19-2.505 GHz). This band covers the frequency bands for Bluetooth (2.4-2.4835 GHz), 2.3 GHz WiMAX (2.3-2.4 GHz) and the lower 2.4 GHz frequency band of WLAN (2.4-2.484 GHz).\\
\indent 
The middle band is centred at 3.5 GHz and has a bandwidth of 430 MHz (3.3-3.73 GHz). Thus, it satisfactorily covers the 3.5 GHz (3.3-3.6 GHz) WiMAX band of operation. The wide band in the 5 GHz range has a centre frequency of 5.5 GHz and a bandwidth of 1890 MHz (4.64-6.53 GHz). It covers the 5.5 GHz (5.25-5.85 GHz) WiMAX as well as the 5.2 GHz (5.15-5.35 GHz), 5.8 GHz (5.725-5.875 GHz) higher frequency WLAN bands of operation.\\
\indent
The xy, yz and xz plane radiation patterns for these three bands have been shown in fig.~\ref{fig9}. It can be seen that the radiation pattern is omni-directional in nature. Besides the peak gains in the 2.3 GHz, 3.5 GHz and 5.5 GHz bands are 1.3dBi, 1.89dBi and 1.96dBi respectively. The gains remain nearly the same throughout their respective band of operation. Table~\ref{table3} summarizes the information associated with the performance of the antenna while Table~\ref{table4} summarizes the information about various bands covered by the antenna. 

\begin{table}[!]
\caption{Summary of antenna performance}
\centering
\begin{tabular}{|l|l|l|l|c|p{2 cm}|}
\hline
  &  {\bf Centre frequency} & {\bf Bandwidth} & {\bf Peak Gain} \\
\hline
{\bf Band 1} & 2.3 GHz & 315 MHz & 1.3 dBi \\
\hline
{\bf Band 2} & 3.5 GHz & 430 MHz & 1.89 dBi \\
\hline
{\bf Band 3} & 5.5 GHz & 1890 MHz & 1.96 dBi \\
\hline
\end{tabular}
\label{table3}
\end{table}

\begin{table}[!]
\caption{Various wireless technologies supported}
\small\addtolength{\tabcolsep}{2pt}
\centering
\begin{tabular}{|l|c|p{3 cm}|}
\hline
{\bf Band 1} &  Bluetooth, 2.3 GHz WiMAX, 2.4 GHz WLAN\\
\hline
{\bf Band 2} & 3.5 GHz WiMAX\\
\hline
{\bf Band 3} & 5.2/5.8 GHz WLAN, 5.5 GHz WiMAX\\
\hline
\end{tabular}
\label{table4}
\end{table}

\section{Comparison with recent antenna designs}
In order to solve some of the performance related issues with existing multiband antenna designs previously described in Section I, some novel simple patch designs have been proposed recently \cite{zhai2013compact,mehdipour2012compact}. In \cite{zhai2013compact}, the patch consists of three circular arc shaped strips and generates three different resonant modes to cover 2.4/5.2/5.8-GHz WLAN and 2.5/3.5/5.5-GHz WiMAX frequency bands of operation.\\
\indent
In case of \cite{mehdipour2012compact}, the antenna makes use of a circular arc shaped patch to achieve a multiband operation to cover wireless applications at 2.4/3.5/5.2/5.8-GHz. These antennas are excellent and indeed overcome some of the fabrication and cost issues due to the simple nature of their design. However, the proposed design has certain benefits over these designs as well.\\
\indent
In \cite{zhai2013compact}, the radiator is around 14 mm x 22.5 mm which is nearly three times in area as compared to that of the proposed design. The bandwidth of the tri-band antenna\cite{zhai2013compact} for the lower, middle and higher bands of operation are 400 MHz (2.38-2.78 GHz), 480 MHz (3.28-3.76 GHz) and 1000 MHz (4.96-5.96 GHz) respectively. Although for the lower and middle band, the bandwidth is slightly higher than that of the proposed design, the additional frequencies does not increase the number of supported wireless applications. Also, in case of the higher band, the frequencies supported by this antenna are also supported by the proposed design. Gains for the lower, middle and higher operation bands in \cite{zhai2013compact} are 2.5 dBi, 2.5 dBi and 3.7 dBi respectively which are higher as compared to design in this paper. In case of \cite{mehdipour2012compact}, the antenna radiator size is 20 mm x 20 mm which is again nearly four times that of the antenna proposed in this paper but has nearly the same bandwidth and peak gains of around 1.2 dBi, 2.6 dBi and 2.3 dBi respectively which are again slightly higher. However, the gains achieved by our design are reasonable and suitable for practical applications. This can be concluded from  \cite{hu2013compact,li2013compact} which mention acceptable gains as low as 0.71 dBi for WLAN and 1.95 dBi for WiMAX. Thus, it can be clearly concluded that in situations wherein a compact design is needed and the associated reductions in gains can be tolerated, the proposed design provides considerable advantage over the existing designs.

\section{Conclusion}
A novel planar antenna with a compact radiator has been proposed and studied in this paper. The proposed antenna resonates at 2.3 GHz, 3.5 GHz and 5.5 GHz with wide bandwidths of 315 MHz (2.19-2.505 GHz), 430 MHz (3.3-3.73 GHz) and 1890 MHz (4.64-6.53 GHz) to cover Bluetooth, 2.3/3.5/5.5 GHz-WiMAX and 2.4/5.2/5.8 GHz-WLAN bands of applications. Main parameters such as the return loss, impedance bandwidth, and far-field characteristics at operating bands have been studied. Therefore, the proposed antenna has satisfactory characteristics and is a promising design for use as a multiband communication antenna. 

\bibliographystyle{unsrt}
\bibliography{main}
\end{document}